\documentclass{amsart}

\usepackage{epsfig}
\usepackage{epsf}
\usepackage{bm}                 
\usepackage{amsmath}
\usepackage{amssymb}
\usepackage[alpine,clock,electronic,geometry,misc,weather]{ifsym}
\usepackage{mathtools}
\usepackage{MnSymbol}
\usepackage{dcolumn}
\usepackage{graphicx}
\usepackage{xcolor}
\usepackage[small,bf, format=plain, width=.75\textwidth, justification=RaggedRight]{caption}[2008/04/01]
\usepackage[
top    = 1.75cm,
bottom = 1.50cm,
left   = 3.00cm,
right  = 2.50cm,
footskip = 1.3cm]{geometry}

\input amssym.def

\makeatletter
\newcommand{\lambdabar}{{\mathchoice
  {\smash@bar\textfont\displaystyle{0.25}{1.2}\lambda}
  {\smash@bar\textfont\textstyle{0.25}{1.2}\lambda}
  {\smash@bar\scriptfont\scriptstyle{0.25}{1.2}\lambda}
  {\smash@bar\scriptscriptfont\scriptscriptstyle{0.25}{1.2}\lambda}
}}
\newcommand{\smash@bar}[4]{%
  \smash{\rlap{\raisebox{-#3\fontdimen5#10}{$\m@th#2\mkern#4mu\mathchar'26$}}}%
}
\makeatother

\begin{document}

\title{Fundamental limits on information processing from black-hole thermodynamics}
\author{Ulvi Yurtsever} \email{ulvi@charter.net}
\address{MathSense Analytics, 1273 Sunny Oaks Circle, Altadena, CA 91001}
\date{\today}

\begin{abstract}
Thought experiments involving thermodynamic principles have often been used to derive
fundamental physical limits to information processing and storage. Similar
thought experiments involving black holes can shed further light into such limits.

\end{abstract}


\maketitle
\thispagestyle{plain}

~~~~~~~~~

Consider the entropy $S( \mathcal R )$ of a system enclosed in a region $\mathcal R$
of spatial size $L$. A classic example of a
fundamental physical limit on computation derived from thermodynamics
and black-hole physics \cite{heb} is an upper limit on $S( \mathcal R )$. It follows from the second
law that $S( \mathcal R )$ cannot exceed \cite{explainBH} the entropy of the largest black hole that would
fit inside $\mathcal R$:
\begin{equation}
S(\mathcal R ) \leqslant S_{\rm max}(\mathcal R )
= k_B \frac{c^3 \mathcal A (\mathcal R )}{4 \hbar \, G}  \equiv S_{BH}(\mathcal R )
\end{equation}
where ${\mathcal A}(\mathcal R ) \approx \pi L^2$ is the surface area enclosing the region. Let us
call the right hand side the ``Bekenstein-Hawking limit" $S_{BH}(\mathcal R )$. Apart from a factor
$1/(k_B \log 2)$, the B$-$H limit is an upper bound (measured in bits) on the amount of information
that can be stored in $\mathcal R$.

{\bf \noindent Landauer entropy from black-hole thermodynamics}

Another well-known physical limit on computation
is the Szilard-Landauer argument \cite{szila} which implies a fundamental
thermodynamic cost for erasing a single bit of information: When a bit
of information is erased, entropy must go up by at least the amount $k_B \log 2$. The argument
employs a ``Maxwell's demon" whose actions are reversible except
for the erasing of information, and who would be able to violate the second law unless
erasure is accompanied with an increase in entropy of at least the Landauer bound.

We can derive a similar lower bound for this
entropy increase from an order-of-magnitude analysis of a thought experiment involving black holes.

Consider a physical system with negligible intrinsic entropy
encoding a single bit of information. The thought experiment
starts with the observation that erasing this information is
physically equivalent to throwing the encoding
system into a black hole.

So what is the minimum amount by which entropy must increase when we throw a single bit into a black hole?

Assume the black hole is Schwarzschild with mass $M$. Its Bekenstein-Hawking entropy is given by
\begin{equation}
S_{bh} = k_B \frac{c^3 A}{4 G \hbar}
\end{equation}
where $A$ is the horizon area $A = 4 \pi R^2$, with the horizon radius
\begin{equation}
R = \frac{2GM}{c^2} \; .
\end{equation}
In terms of mass, the black-hole's entropy is then
\begin{equation}
S_{bh} = k_B \frac{4 \pi G}{ c \hbar} M^2 \;.
\end{equation}
For the system encoding the single physical bit to fall completely into the black hole, the Heisenberg
uncertainty relation imposes a fundamental lower limit on its energy: the de Broglie wavelength
of the system (given in terms of its momentum $p$) must be smaller than the effective
horizon diameter:
\begin{equation}
\lambdabar = \frac{\hbar}{p} \lessapprox 2R \mu
\end{equation}
where $\mu$ is a dimensionless geometric
constant of order unity such that the effective capture
cross section of the black hole is $\pi {\mu}^2 R^2$.
Typically we would expect $\mu > 1$ (for example,
for a Schwarzschild hole $\mu = \sqrt{27/4}$
for capturing relativistic particles ~\cite{capture}).
Denoting the energy of the system by $\delta e$ and its rest mass by $\delta m_0$, we have
\begin{equation}
(\delta e)^2 = p^2 c^2 + (\delta m_0 )^2 c^4 \; ,
\end{equation}
from which it follows that
\begin{equation}
p = \sqrt{\frac{{(\delta e)}^2}{c^2} - {(\delta m_0 )}^2 c^2 } \leqslant \frac{\delta e}{c} \; .
\end{equation}
Combining Eq.\,(7) with Eq.\,(5) gives
\begin{equation}
\frac{\delta e}{c} \geqslant p \gtrapprox \frac{\hbar}{2R \mu} \; .
\end{equation}
When the physical bit falls into the black hole, the hole's mass must increase by the amount
$\delta M = \delta e / c^2$.
According to Eq.\,(4), the increase in the black-hole's entropy due to an increase $\delta M$ in
its mass is given by
\begin{equation}
\delta S_{bh} = k_B \frac{8 \pi G}{c \hbar} M \, \delta M \;.
\end{equation}
Now substitute $\delta M = \delta e /c^2$ in Eq.\,(9) and use inequality (8) to obtain
\begin{equation}
\delta S_{bh} \gtrapprox k_B \frac{8 \pi G}{c \hbar} M \frac{\hbar}{2 c R \mu} \;,
\end{equation}
which, when combined with Eq.\,(3) gives
\begin{equation}
\delta S_{bh} \gtrapprox \frac{2 \pi}{\mu} k_B \; .
\end{equation}
Given the qualitative
nature of the present analysis, Eq.\,(11) is in order-of-magnitude
agreement with the more quantitatively rigorous
Szilard-Landauer argument ~\cite{extremal1}.

{\bf \noindent A new bound on information storage}

Let us now turn the argument around, and accept the Landauer bound as a given thermodynamic constraint
on computation, a consequence of the second law deduced from the usual Maxwell's demon construction. Consider
a physical system $\mathcal C$
encoding $n$ bits of information. Let the system have total energy $U$ (including
any rest-mass or thermal energy) and intrinsic thermodynamic entropy $S$. If this system falls into a black
hole of mass $M$, the final entropy of the black hole must be larger than the initial entropy of
the combined system $\mathcal C$ and the black-hole by at least the Landauer bound
\begin{equation}
n k_B \log 2 \; .
\end{equation}
We can write this statement in the form of an inequality as follows:
\begin{equation}
{S_{bh}}^{\rm after} \geqslant {S_{bh}}^{\rm before} + S + n k_B \log 2 \;.
\end{equation}
The initial entropy ${S_{bh}}^{\rm before}$
of the black hole is given by Eq.\,(4), while its final entropy is
\begin{equation}
{S_{bh}}^{\rm after} = k_B \frac{4\pi G}{c \hbar} \left( M+\frac{U}{c^2}\right)^2 
= k_B \frac{4 \pi G}{c \hbar} M^2 + k_B \frac{8\pi G}{c^3 \hbar} M \, U + k_B \frac{4 \pi G}{c^5 \hbar} U^2 \; \;.
\end{equation}
Substituting Eq.\,(14) in the inequality Eq.\,(13) we obtain
\begin{equation}
n k_B \log 2 \leqslant k_B \frac{4\pi G}{c^5 \hbar} U^2 - S +
k_B \frac{8\pi G}{c^3 \hbar} M \, U \; . 
\end{equation}
In the inequality Eq.\,(15), the first three terms are independent of the black-hole mass $M$ but
the last term depends on it. Consequently, Eq.\,(15) represents a family of bounds on $n$ depending
on the parameter $M$. Clearly, the tightest bound obtains with the smallest possible mass
$M_{\rm min}$. If the
system $\mathcal C$ has spatial extent $L$, the smallest black hole that can absorb $\mathcal C$
has horizon diameter equal to $L$; therefore
\begin{equation}
\frac{4 G M_{\rm min}}{c^2} = L
\end{equation}
and we can replace $M$ in Eq.\,(15) by $M_{\rm min}$ to obtain the final inequality
\begin{equation}
n \log 2 \leqslant \frac{4 \pi G}{c^5 \hbar} U^2 - \frac{S}{k_B} +
 \frac{2 \pi}{c \hbar} L \, U \; . 
\end{equation}
In Eq.\,(17) we have a new upper bound ~\cite{extremal2} on the amount of information $n$ (in bits)
that can be stored in a system $\mathcal C$ of total energy $U$, intrinsic entropy $S$,
and spatial extent $L$.

The inequality Eq.\,(17) in general gives much tighter bounds on
information storage than the holographic Bekenstein-Hawking bound Eq.\,(1). As an example,
take $\mathcal C$ to be an everyday computing device with $L \approx 0.1\rm m$ and
rest mass $\approx 1\rm kg$. The B$-$H bound on maximum information storage in this device would
yield a number on the order of $c^3 L^2 / (\hbar G) \sim
4 \times 10^{67}$ bits. In the bound Eq.\,(17), $U$
is dominated by the rest-mass term $(1 {\rm kg}) c^2 \approx 10^{17} \rm J$, the first term
on the right side is
on the order $10^{16}$, and the second is on the order $10^{23}$ (Avogadro's number). The
third and dominant term is approximately $2 \times 10^{42}$ bits, 25 orders of magnitude smaller
than the B-H bound (though still far from limits imposed by current practical considerations).

As another interesting example, let us calculate the maximum
amount of information that can be stored in a
single femtosecond optical pulse \cite{femto}. Consider an optical pulse
of 10fs duration and 1GW pulse power. The total energy of the pulse
is then $U \sim 10 \mu{\rm J} = 10^{-5} \rm J$. Assuming the intrinsic entropy $S$
of the pulse is negligible, the first two terms on
the right hand side of Eq.\,(17) can again be neglected,
and the third term, with $L = 10^{-14} {\rm s} \times c \sim 10^{-6} \rm m$,
yields $(2 \pi / (c \hbar)) L U \sim 2 \times 10^{15} {\rm bits} \sim 200 \;\rm Terabytes$,
a bound relevant enough to current real-world technology to be intriguing.
The Bekenstein-Hawking
bound for this system, on the other hand, would give an upper limit
$c^3 L^2 / (\hbar G) \sim 4 \times 10^{57} \; \rm bits$,
42 orders of magnitude further away from the new bound.

%
%
%
%

\renewcommand{\thefootnote}
{\ensuremath{\fnsymbol{footnote}}}
\addtocounter{footnote}{2}


\begin{thebibliography}{5}
\bibitem{explainBH} Since otherwise one can let the contents of $\mathcal R$ collapse gravitationally
into a black hole and decrease total entropy.
\bibitem{extremal1} By carrying out the same analysis with a close-to-extremal Kerr-Newman
(instead of a Schwarzschild) black hole, it is possible to obtain the sharper inequality
$\delta S_{bh} \gtrapprox ({\pi}/ \mu ) k_B$ instead of  Eq.\,(11).
\bibitem{extremal2} Once again the geometric coefficients in the inequality Eq.\,(17) can be sharpened
(replaced by smaller coefficients of order unity)
by considering close-to-extremal black holes instead of the Schwarzschild Eq.\,(4).
\end{thebibliography}
\end{document}